\documentclass{elsart}
\usepackage{epsfig}

\begin{document}

\begin{frontmatter}

\title{Phase Synchronization in Temperature and Precipitation Records}

\author[adr1,adr2]{Diego Rybski \corauthref{cor1}},
\author[adr1]{Shlomo Havlin},
\author[adr2]{Armin Bunde}
\corauth[cor1]{Corresponding author. Institut f\"{u}r Theoretische Physik III, 
Universit\"{a}t Giessen, Heinrich-Buff-Ring 16, D-35392 Giessen, Germany.
E-mail: Diego.Rybski@physik.uni-giessen.de (D. Rybski).}
\address[adr1]{Minerva Center and Department of Physics, Bar Ilan University, 
Israel}
\address[adr2]{Institut f\"{u}r Theoretische Physik III, Universit\"{a}t 
Giessen, D-35392 Giessen, Germany}

\begin{abstract}\label{abstract}
We study phase synchronization between atmospheric variables such as daily mean 
temperature and daily precipitation records. We find significant phase 
synchronization between records of Oxford and Vienna as well as between the 
records of precipitation and temperature in each city. To find the time delay 
in the synchronization between the records we study the time lag phase 
synchronization when the records are shifted by a variable time interval of 
days. We also compare the results of the method with the classical 
cross-correlation method and find that in certain cases the phase 
synchronization yields more significant results.
\end{abstract}

\end{frontmatter}

\section{Introduction}\label{intro}
In recent years there was much interest in long term persistence of temperature 
records \cite{EVA2} detected by Detrended Fluctuation Analysis \cite{Peng}.
Fluctuations in space and time of meteorologic records are usually 
characterized by Teleconnection Patterns \cite{Barnston1987}. They describe 
recurring and persistent patterns of circulation anomalies that take place in 
huge geographical domains. Prominent patterns are the North Atlantic 
Oscillation (NAO) that appears all over the year, or the East Atlantic Pattern 
(EA), which appears from September to April. Each site in the Teleconnection 
Patterns is characterized by the strength of the cross-correlation of this site 
with all other sites in the pattern, for a given meteorologic parameter. By 
this, a correlation matrix is defined, which usually exhibits two to four 
regions of extreme high or low values, which are called "centers of action". A 
more recent measure is the Rotated Principal Component Analysis (RCPA), which 
uses eigenvectors of the correlation (or cross-covariance) matrix after certain 
scaling and rotation to identify the meteorologic patterns.

The methods to identify teleconnection patterns are based on cross-correlation 
which essentially compares the amplitude records. Here we suggest an 
alternative method for studying relations between meteorological records, which 
is based on the phase synchronization approach \cite{Rosenblum1996}. We show 
that this method can be applied also to complex signals where the fluctuations 
are not pure oscillations. For certain meteorological records, we find that the 
phase synchronization approach performs better than the conventional 
cross-correlation approach. The method also enables to quantify the typical 
wavelengths of a signal, which cannot be detected by cross-correlation.\par

The paper is organized as follows: We describe the phase synchronization method 
in Section \ref{method}, present the results in Section \ref{results} and 
discuss and summarize them in Section \ref{discussion}.

\section{Phase Synchronization Method}\label{method}
The Phase Synchronization Method was originally applied to weakly coupled 
chaotic oscillators. The method enables to reveal relations between two 
complex records by focusing on the phases of the fluctuations in each record. 
The technique was found very useful for identifying phase synchronization in 
several biological systems, including the synchronization between the breathing 
cycle and the heart rhythm \cite{Schaefer}, which reveals the weak interaction 
between the human respiratory and the cardiovascular system. Analysis of  
synchronization has also been performed in ecological systems, where complex 
population oscillations occur \cite{Blasius}. For more applications and a 
pedagogical review of the method we refer to \cite{KurthsBook}.

Generally, two periodic oscillators are in resonance, if their frequencies 
$\omega_1$ and $\omega_2$ are related by
\begin{equation}\label{resonance}
n\omega_1\approx m\omega_2\quad,
\end{equation}
where $n$,$m$ are integers. 
We define a phase $\phi_j(t) = \omega_j t$ for each oscillator, and the 
generalized phase-difference is $\varphi_{n,m}=n\phi_1(t)-m\phi_2(t)$. Hence we 
have resonance for the condition
\begin{equation}\label{phasediff}
|\varphi_{n,m}-\delta|<\mbox{const.}\quad,
\end{equation}
where $\delta$ represents the phase shift between both oscillators, and the 
constant on the r.h.s. is any positive finite number. This condition holds also 
when the frequencies are fluctuating. In this case, $\phi_j(t)$ is calculated 
for each single record by using a Hilbert transform (see below). In order to 
test for phase synchronization, we determine \cite{Rosenblum2001} 
\begin{equation}\label{relphase}
\psi_{n,m}=\varphi_{n,m}\mbox{ mod }2\pi\quad.
\end{equation}
If the histogram of $\psi_{n,m}$ shows a maximum at a certain phase-difference, 
the two records are synchronized at this phase.\\
In practice, the phase synchronization analysis of two records of length $N$ 
consists of five steps:
\begin{itemize}

\item{In the first step, we construct from the scalar signals $\tau_{j}(t)$, 
$j=1,2$, the complex signals $\zeta_j(t)=\tau_j(t)+i\tau_{H_j}(t)=
A_j(t)e^{i\phi_j(t)}$, where  $\tau_{H_{j}}(t)$ is the Hilbert transform of 
$\tau_j(t)$ \cite{Gabor}.}
\item{Then we extract the phases 
$\phi_{1}(t)$ and $\phi_{2}(t)$.}
\item{Next we cumulate the phases such that every cycle, the phases 
$\phi_{j}(t)$ increase by $2\pi$.}
\item{Then we quantify the difference of the phases 
$\varphi_{n,m}(t)=n\phi_1(t) - m\phi_2(t)$.}
\item{Finally, we create a histogram of $\psi_{n,m}=
\varphi_{n,m}\mbox{ mod }2\pi$ for various $m$ and $n$ values. To do this, we 
subdivide the possible range of the phases
$\psi_{n,m}$ into $M$ intervals (bins) of size
$2\pi/M$ and determine how often the phase $\psi_{n,m}$ occurs in each 
interval.}
\end{itemize}

In the absence of phase synchronization, the histogram of $\psi_{n,m}$ is
expected to be uniform, because all phase-differences occur with the same 
probability. In the presence of phase synchronization, there exists, for a 
certain pair $(m,n)$ a peak in the histogram.\par

To quantify the significance of synchronization we use an index 
\cite{Rosenblum2001} based on the Shannon entropy $S$:
\begin{equation}\label{phasesyncindex}
\rho_{n,m}=\frac{S_{max}-S}{S_{max}}\quad, 
\end{equation}
where $S=-\sum_{k=1}^{M}p_k\ln p_k$ and $p_k$ is the probability of finding 
$\psi_{n,m}$ in the $k$-th bin of the histogram. By definition, the maximum 
entropy is $S_{\rm max}=\ln M$. The synchronization index is restricted to the 
unit interval $0\le\rho_{n,m}\le1$ and is minimal for a uniform distribution 
and maximal in the case of a $\delta$-function.\par

By introducing a time lag into the phase synchronization method, realized by 
a certain shifting interval $S$ between the two records, it may occur, that for 
some cases best phase synchronization is found only for a certain time lag. In 
this case, the synchronization is delayed by this interval, which can be
determined by the position of the peak in the synchronization index.

\section{Results}\label{results}
We begin the demonstration of the method on the temperature and precipitation 
records of Oxford (GBR) and Vienna (AUT). The stations are distant enough in 
order not to give trivial results, but of sufficient closeness for their 
climate to interact. In order to analyze only the fluctuations, we deseasoned 
the records by subtracting the annual cycles. We mostly discuss the temperature 
time series. The values for the first hundred days of the year 1873 for both
cities are shown in Fig. 1. Although some similarities can be 
guessed, the question how to quantify these similarities is of interest. One 
method is the cross-correlation approach. Here we propose that complementary 
information can be revealed by the phase synchronization method.\par

Figs. 2 and 3 demonstrate the steps of the method. In 
Fig. 2(a) a small section of the temperature record measured at 
Oxford is given. The corresponding phases (Fig. 2(b)) were 
determined using Hilbert transform. Fig. 2(c) shows what the cumulated 
phases look like, where after every cycle $2\pi$ is added. In Fig. 3(a) the 
cumulated phases for both complete records are shown, while 
in Fig. 3(b) the phase-differences are displayed. The histogram of 
the phase-differences modulo $2\pi$, is given in Fig. 3(c). A clear peak can 
be seen in the histogram and the synchronization index is $\rho=0.0242$. This 
value can not be improved by taking n:m other than 1:1.\par

To get information about the significance of this result we perform time lag
phase synchronization, i.e. a shifting-test, where the series are shifted 
against each other by a given interval of $S$ days. The non-overlapping values 
in both sequences are ignored for the process. Obviously in the case of no 
synchronization the value of $\rho$ must be lower than in synchronization. In 
Fig. 4(a) the result of shifting is shown. For shifting of 
several days in both directions the synchronization decreases dramatically. A 
histogram for a shift of +20 days, where $\rho=0.0011$, is given in Fig. 
4(c). This shifting-test reveals, that the non-shifted case does 
not correspond to the best synchronization. A higher synchronization index 
($\rho=0.0315$) can be achieved with a shift of -1 day. Fig. 
4(b) shows this histogram. The peak is slightly sharper and 
higher than in Fig. 3(c). This result is reasonable since a cycle of 
fluctuation which is detected in Oxford reaches Vienna (due to high latitude 
western winds) about one day later.\par

The results are less pronounced for the two precipitation time records. The 
shifting-test is displayed in figure 5(a). Here the importance of 
the shifting-test becomes clear. Even when shifted, $\rho$-values of the order 
of $0.004$ are achieved, but the peak is still significantly higher. This high 
background is probably due to the large fluctuations of the precipitation 
records. They show a spiky structure, that leads the Hilbert transform to give 
many slips and phases of short duration. Indeed, apart from the long cycles, 
these records consist of many of 3 to 4-day-periods (shown in Fig. 6(b)). 
When a pair of precipitation records is shifted, the phases 
still show matching because of the multitude of very short periods, yielding 
noise-induced synchronization in the background. Nevertheless a dominant peak 
is obtained in this representation, the maximum synchronization with 
$\rho=0.0072$ is reached when the series are shifted by -2 days. Note that this 
value is a factor $4$ smaller than that for temperature.\\
Synchronization is also found between temperature and precipitation records at 
the same site. In Oxford the temperature and precipitation records are very 
weakly synchronized (Fig. 5(b)), with a small peak of 
$\rho=0.0009$. At Vienna (Fig. 5(c)) the peak is at least six 
times larger. In both cases the peaks are located at a time lag of +1 day, 
i.e., they are better synchronized when the temperature record is one day in 
advance to the precipitation record. In comparison to synchronization of the 
two precipitation time series, they have much less noise-induced 
synchronization in the background.\\
The fact that best synchronization for temperature records of Oxford and Vienna 
is found when they are shifted by one day, exhibits the statistical delay 
between cycles of weather at both sites. This conclusion is supported by the 
result for precipitation series, where the maximum $\rho$ occurs for shifting 
of two days. Probably the real delay is approximately 1.5 days, which, due 
to low sample-rate, can not be determined more precisely.\par

Usually Fourier Transform is applied in order to discover dominant global 
frequencies or wavelengths in a considered time series. But no direct 
information can be gained about cycles of varying wavelength, since Fourier 
Transform detects global waves in the record. We suggest to use the cumulated 
phases to estimate the wavelengths in the time series. Namely, we count the 
days, until the phases pass steps of $2\pi$, and generate a histogram with 
frequency of occurrence versus wavelength.\\
For the temperature and precipitation records of Oxford these histograms are 
shown in Fig. 6. In the case of temperature (Fig. 6(a)) the fluctuations have 
a wide range from $2$ to about $90$ days, but most of them take five to ten 
days. The precipitation record (Fig. 6(b)) consists of much more fluctuations 
of short wavelength. The length of three days occurs $274$ times, while only 
$84$ times in the temperature record. Also the maximum wavelength of the 
precipitation record is only about $50$ days. Note that the considered cycles 
are not periodic, but have random wavelengths.

\section{Discussion}\label{discussion}
Comparing phase synchronization with the classical cross-correlation method is 
of interest. While in phase synchronization the phases of the cycles play the 
major role and not the amplitudes, in cross-correlation both aspects are 
superimposed. Thus we expect to obtain complementary information from the two 
approaches. In the following two examples more significant results were 
obtained from phase synchronization compared to cross-correlation.\\
In Fig. 7 we compare time lag phase synchronization and 
time lag cross-correlation for precipitation in two sites in Asia. The phase 
synchronization index exhibits a distinct peak with a maximum at $-3$ days,
while the cross-correlation only gives large background noise with a peak that 
is almost indistinguishable from the background.
Fig. 8 also demonstrates an advantage of phase
synchronization. It compares phase synchronization and cross-correlation for 
records without annual deseasoning. Here only the average value of each record 
was subtracted. It is seen, that while in the phase synchronization almost a 
constant background with $\rho\approx0.19$ is obtained, the cross-correlation 
shows large annual oscillations, as expected. Thus, the peak in phase 
synchronization ($\rho=0.25$) compared to the background (Fig. 8(a)) is much 
more significant than that in cross-correlation analysis (Fig. 8(b)). The high 
value of the constant background in the time lag phase synchronization is due 
to the annual synchronization, which is almost not influenced by variation of 
the time lag. The peak in the time lag phase synchronization (Fig. 8(a)) is 
thus mainly due to phase synchronization of the fluctuations which represents 
irregular cycles of deviation from the mean annual cycle.\\
Synchronization in the atmosphere plays an important role in climatology. 
For example tests on an atmospheric global circulation model have been done, 
where the complete synchronization of the two hemispheres has been analyzed 
\cite{Lunkeit}. What does phase synchronization in climate records mean? 
For temperature records, e.g. a complete relatively warm period followed by a
cold period represents a cycle in terms of phases. At another site, which is 
synchronized to the first, statistically a similar cycle also occurs, maybe 
with some delay. The amplitudes of these cycles have no influence on the phase 
synchronization. This is in contrast to cross-correlation, which is strongly 
affected by the amplitudes. Thus, phase synchronization might be useful when 
interaction in records of different climate regions is analyzed, such as 
maritime, where temperature fluctuations are less pronounced, and continental 
regions with larger fluctuations.

\section*{Acknowledgments}
We are grateful to Prof. Dr. H.-J. Schellnhuber and Dr. H. \"Osterle from the 
Potsdam Institute for Climate Impact Research (PIK) for providing the
temperature and precipitation records as part of a joint research cooperation. 
Further we wish to thank Prof. Steve Brenner for discussions on 
Teleconnections. 
We like to acknowledge financial support by the Deutsche Forschungsgemeinschaft
and the Israel Science Foundation.

\begin{figure}\begin{center}
\epsfxsize=11cm	
\epsffile{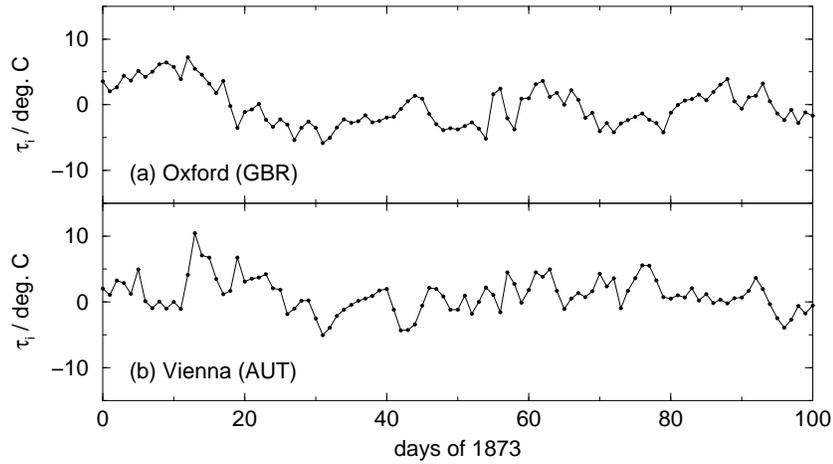}\end{center}
\caption{A typical example of daily mean temperature record for 100 days, 
starting in 1873 at (a) Oxford and (b) Vienna, after subtraction of the annual 
cycles, average over all years of the record.}
\end{figure}

\begin{figure}\begin{center}
\epsfxsize=11cm	
\epsffile{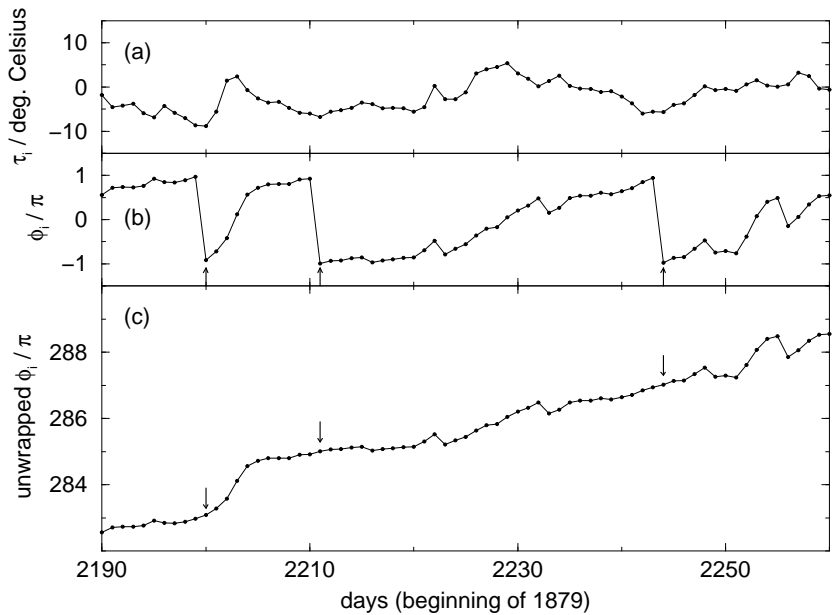}\end{center}
\caption{The steps from the signal to the cumulated phases. (a) Part of 
the deseasoned temperature record measured at Oxford. (b) The phases extracted 
by the Hilbert transform. (c) The cumulated phases. The arrows represent the 
edges of the cycles.}
\end{figure}

\begin{figure}\begin{center}
\epsfxsize=\textwidth	
\epsffile{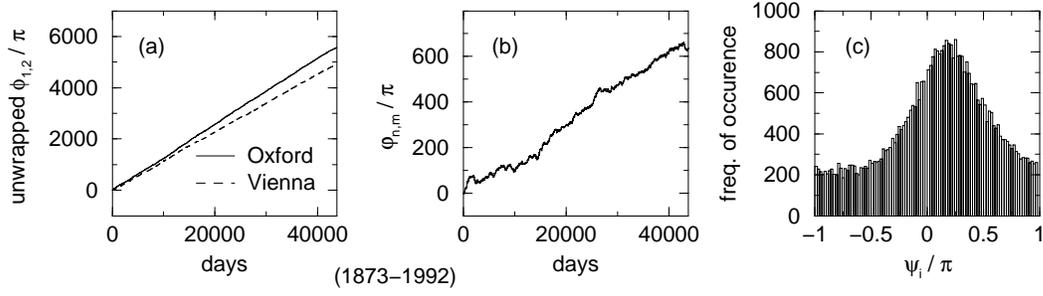}\end{center}
\caption{(a) Cumulated phases for the deseasoned temperature record of 
Oxford (solid line) and Vienna (dashed line) for the years from 1873 to 1992. 
(b) Phase-difference. (c) Histogram of phase-difference mod $2\pi$ (100 bins).}
\end{figure}

\begin{figure}\begin{center}
\epsfxsize=\textwidth	
\epsffile{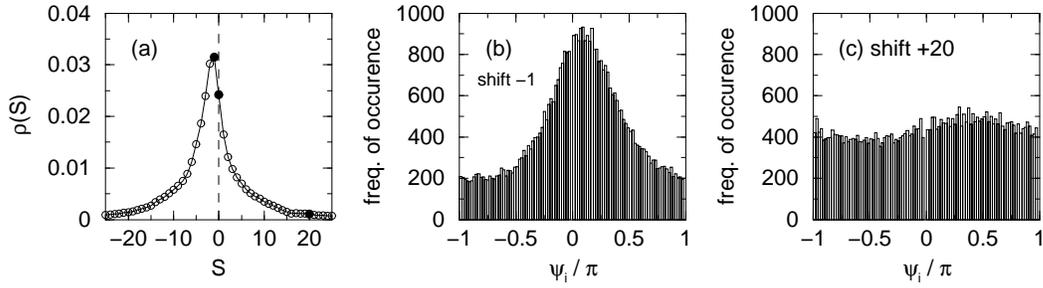}\end{center}
\caption{(a) Dependence of the synchronization index $\rho$ on the 
shifting interval $S$ between the records of Oxford and Vienna. Negative 
shifting means that the series of Vienna is in advance. The corresponding 
non-overlapping values were cut from the records. Histograms for the filled 
dots are shown in (b) with a shift of -1, (c) with shift of +20. Note that in 
Fig. 3(c) the records are not shifted.}
\end{figure}

\begin{figure}\begin{center}
\epsfxsize=12cm	
\epsffile{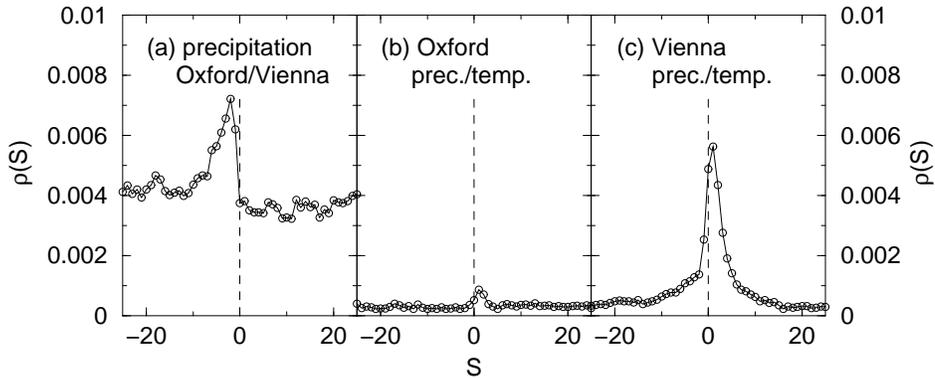}\end{center}
\caption{The synchronization index $\rho$ as a function of the shifting 
interval $S$ for different pairs of daily records. (a) Precipitation at Oxford 
and Vienna (1873-1989), whereas for negative shifting values we cut the 
beginning of Vienna's and the end of Oxford's record. Synchronization between 
temperature and precipitation measured at (b) Oxford (1873-1992) and (c) Vienna 
(1873-1989).}
\end{figure}

\begin{figure}\begin{center}
\epsfxsize=8.6cm
\epsffile{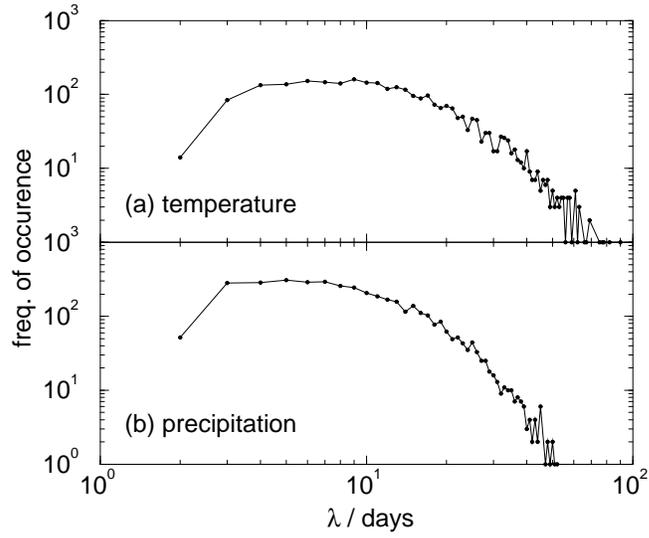}\end{center}
\caption{Histogram of wavelengths for (a) temperature and (b) 
precipitation records at Oxford, determined by counting the number of days for 
which the phases complete a cycle.}
\end{figure}

\begin{figure}\begin{center}
\epsfxsize=12cm	
\epsffile{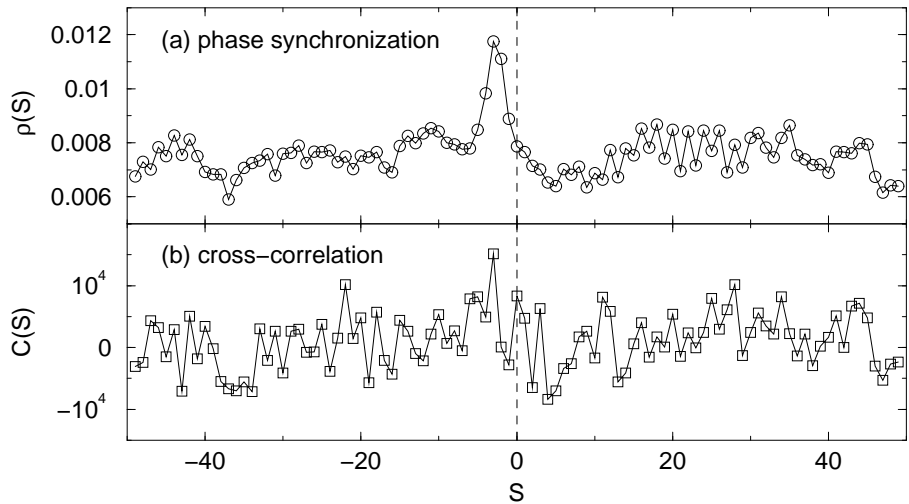}\end{center}
\caption{An example for comparison between (a) phase synchronization and 
(b) cross-correlation. The methods were applied to the precipitation records of 
Wulumuqi (CHN) and Pusan (KOR) (daily 1951-1990). Negative shifting means
that the dates of Pusan correspond to earlier dates of Wulumuqi.}
\end{figure}

\begin{figure}\begin{center}
\epsfxsize=12cm	
\epsffile{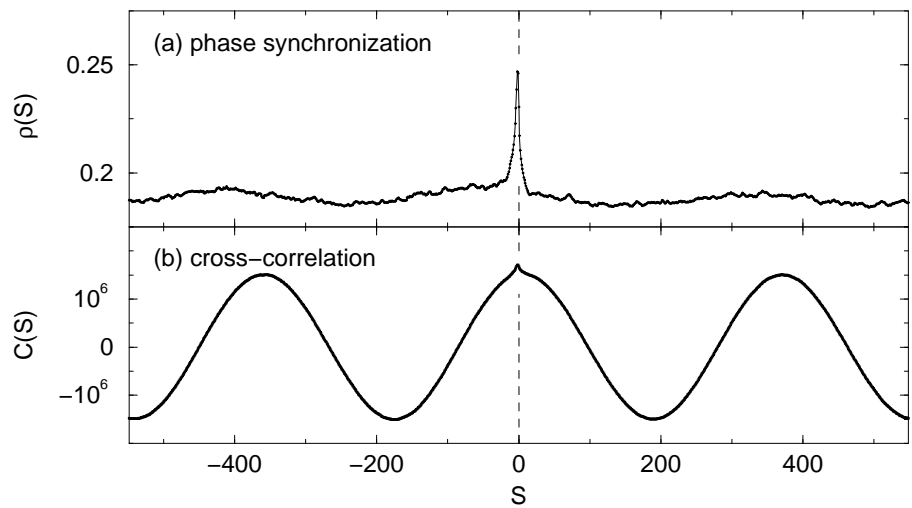}\end{center}
\caption{Comparison of (a) phase synchronization and (b) cross-correlation.
Again we analyzed the daily mean temperature records of Oxford and Vienna
(1872-1992), but here we subtracted the global average from the record instead 
of the annual cycle before performing the methods. For positive $S$ the record 
of Oxford is in advance. Note that (a) corresponds to Fig. 
4(a).}
\end{figure}

\end{document}